\documentclass{article}
\usepackage{spconf,amsmath,graphicx,hyperref,booktabs,amsfonts,amssymb}

\title{On Temporal Binding in Large Audio Language Models}

\name{Paul Primus$^{1}$ and
      Gerhard Widmer$^{1,2}$}
\address{$^1$Institute of Computational Perception, $^2$LIT Artificial Intelligence Lab \\ Johannes Kepler University Linz, Austria}
\begin{document}
\maketitle

\begin{abstract}
Reasoning about temporal structure of audio recordings requires Large Audio Language Models (LALMs) to associate sound events with their temporal position. Understanding the underlying mechanisms is a first step toward diagnosing failures and identifying model components that may need improvement. Using mechanistic interpretability, we investigate how temporal information is represented and bound to sound events in three open-source LALMs. We find that across all three, event-specific location becomes concentrated in event name representations at intermediate modality integration layers. These representations encode coarse event position along a low-dimensional, curved relative time trajectory. Steering event name representations along this trajectory systematically shifts before/after beliefs, providing evidence that these representations contribute to coarse temporal reasoning. In contrast, the same interventions do not reliably shift predicted onset timestamps, suggesting that coarse temporal reasoning and precise metric event localization rely on distinct mechanisms.
\end{abstract}


% ============================================================================
\section{Introduction}
\label{sec:intro}
% ============================================================================
Temporal reasoning over an audio recording requires Large Audio Language Models (LALMs) to associate sound events with their corresponding temporal properties. 
Determining whether footsteps occurred before or after a vacuum cleaner, for example, requires recognizing both events, associating each with a coarse temporal position, and comparing these event-bound representations. 
More fine-grained localization of events would additionally require identifying event boundaries and mapping them onto a continuous time axis.
Recent LALMs have demonstrated progress on temporal audio reasoning benchmarks, performing well on coarse reasoning tasks while still exhibiting notable shortcomings \cite{ghosh2024compa,bhattacharya2025trea,sakshi2025mmau,kulkarni2026closerlook}. 
Despite the progress, it remains unclear how LALMs internally bind temporal positions to events and which hidden activations are primarily involved in temporal reasoning.
Understanding these internal mechanisms can help localize the source of temporal reasoning failures and indicate which model components may need improvement.
%Current LALMs perform remarkably well on temporal reasoning benchmarks \cite{ghosh2024compa,bhattacharya2025trea}; however, it remains unclear how they internally bind temporal positions to events and which hidden activations are primarily involved in temporal reasoning. 
%Understanding these internal mechanisms can help localize the source of temporal reasoning failures and indicate which model components may need improvement.
Mechanistic interpretability methods, such as probing \cite{alain2017probes}, causal tracing \cite{meng2022locating}, activation patching \cite{zhang2024patching}, and representation steering \cite{li2023inference}, provide tools for this by revealing where information is encoded and how it causally affects model behavior.
Recent studies of LALMs have used mechanistic interventions to investigate audio--text fusion \cite{chen2026audiotextfusion} and identify audio-specialized attention heads that strengthen audio grounding \cite{glazer2026audiospecialists}.
% , and redirect temporal attention toward queried sound events \cite{lin2026steering}.
Kang et al. \cite{kang2026spatiotemporal} identified low-dimensional spatiotemporal representations in vision- and video-language models, showing that location information becomes bound to textual object activations and causally influences model predictions.
We investigate whether analogous event-bound temporal representations emerge in LALMs and whether they support both coarse temporal reasoning and precise event localization. We study AF-Next \cite{ghosh2026afnext}, MOSS-Audio \cite{yang2026mossaudio}, and Qwen3-Omni \cite{xu2025qwen3omni} using activation swapping and representation analysis to characterize these representations and test their causal role. Our results suggest a common mechanism for coarse temporal reasoning, while precise localization appears to rely on a separate mechanism.

\begin{figure*}[ht!]
    \centering
    \includegraphics[width=\linewidth]{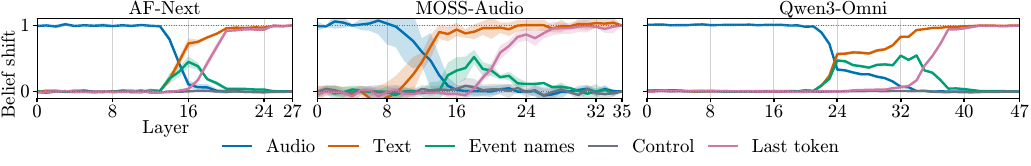}
    \caption{Activation swapping across layers for three audio language models. Each line shows the belief shift $S_L$ when residual stream activations for a specific token group are replaced with activations from a second forward pass using the same text query but audio with the event order reversed. Values near 1 indicate a strong shift, while values near 0 indicate little effect.}
    \label{fig:swap}
\end{figure*}

\section{Experimental Setup}
\label{sec:setup}

For our controlled experiments, we construct synthetic sound scenes from ESC-50 foreground events \cite{piczak2015esc50} and FreeSound backgrounds \cite{font2013freesound}. 
Foreground clips are trimmed of leading and trailing silence and mixed into the backgrounds at SNRs between approximately $-5$ and $0\,\mathrm{dB}$. 
For each background, ESC-50 classes detected by an audio tagger are excluded from insertion to reduce overlap with preexisting events. 
We construct disjoint development and test sets for the synthetic experiments.
We additionally use subsets of RealDESED \cite{schmid2026realdesed} for validating our experiments on natural audio.
RealDESED is well suited for this purpose, as it contains longer audio recordings of 15--30\,s duration with strong temporal annotations for 15 sound event classes. Depending on the experiment, we select recordings containing either two clearly ordered events suitable for before/after reasoning or a clearly identifiable event with limited temporal overlap with other events. Code and details for reproducing our experiments are available at \url{https://github.com/OptimusPrimus/icassp2027_temporal_binding}.

% ============================================================================
\section{Temporal Information Is Bound to Event Name Representations}
\label{sec:binding}
% ============================================================================

We first identify where event-specific temporal information enters the text residual stream and then test how precisely event position can be decoded from the representations.

\subsection{Intermediate Layers Bind Temporal Information to Event Name Tokens}
\label{sec:swap}

Following Kang et al. \cite{kang2026spatiotemporal}, we ask where event-specific temporal information enters textual representations and causally contributes to temporal reasoning. 
Using our synthetic dataset, we construct pairs of 10\,s recordings containing the same background and two ESC-50 events in mirrored temporal arrangements. 
Let $c_q$ and $c_r$ denote the query and reference events. 
If recording $x$ contains $c_q$ before $c_r$, its paired recording $y$ contains the same events in the opposite order. 
Both recordings are paired with the identical prompt: ``Does [query event] occur before or after [reference event]?'' 
At each layer, we replace subsets of residual stream activations for $x$ with the corresponding activations from $y$ and continue the forward pass. 
If the replaced activations contribute event-specific temporal information, the intervention should shift the model's output from its belief for $x$ toward its belief for $y$.
We quantify this using a normalized \emph{belief shift} $S_L$.
Under teacher forcing, we condition on the response prefix ``[query event] occurs'' and measure the probabilities of ``before'' and ``after''. 
Let $a^\star \in \{\textrm{before},\textrm{after}\}$ denote the ground truth relation in $x$, and let $p_z(a^\star)$ denote the probability assigned to this answer for input $z$. 
For an intervention at layer $L$,
\begin{equation}
    S_L =
    \frac{
        p_x(a^\star)-p_{\widetilde{x}_L}(a^\star)
    }{
        p_x(a^\star)-p_y(a^\star)
    },
    \label{eq:belief_shift}
\end{equation}
where $\widetilde{x}_L$ denotes the intervened forward pass. 
Thus, $S_L=0$ indicates no effect, whereas $S_L=1$ reproduces the full belief change induced by the mirrored recording.

Figure~\ref{fig:swap} compares interventions for selected token groups. 
Across compared models, a consistent information flow pattern emerges: In early layers, swapping the audio representations (blue) strongly affects the model's temporal belief, whereas swapping all text tokens (orange) has little effect. 
In an intermediate layer range, the audio influence decreases while swapping only the event name tokens (green) produces a substantial belief shift. 
At later layers, the event name effect fades and interventions on the full textual sequence become dominant, with interventions on the final token (pink) eventually having the strongest impact of all individual tokens. 
The ``before or after'' control tokens (grey) remain close to zero.
These results suggest that event-specific temporal information becomes bound to textual representations in intermediate event name activations; swapping these representations shifts the predictions, which indicates that they are causally involved in temporal reasoning.

\begin{figure*}[ht!]
    \centering
    \includegraphics[width=\linewidth]{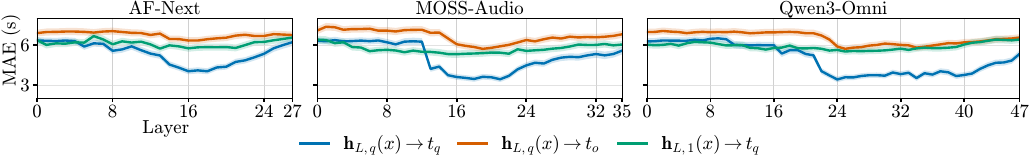}
    \caption{Layerwise MAE for decoding query- and unmentioned-event midpoints ($t_q$, $t_o$) from token activations $\mathbf{h}_{L,q}$ and $\mathbf{h}_{L,1}$.}
    \label{fig:decode}
\end{figure*}

\subsection{Coarse Temporal Position Is Decodable from Event Name Representations}
\label{sec:decoding}

We next test how precisely an event's temporal position can be recovered from these event name representations. 
To this end, we construct 30\,s synthetic recordings containing two distinct ESC-50 events at random temporal locations. 
One event is selected as the query event, and the model is prompted with ``Is there [query event]?''.
Let $\mathbf{h}_{L,q}(x)\in\mathbb{R}^d$ denote the layer-$L$ residual stream activation of the final token of queried event $c_q$, and $\mathbf{h}_{L,1}(x)$ that of the first text prompt token, i.e., the one corresponding to ``Is''.
We train a MLP regressor with one hidden layer to predict event midpoints of the query and the unmentioned event ($t_q$ and $t_o$) from these activations at each layer on the synthetic training set. 
To test whether the decoded temporal signal generalizes across event identity, we use a five-fold cross-validation, training on 40 query classes and evaluating on the remaining 10.

Figure~\ref{fig:decode} shows that the queried event midpoint $t_q$ is most accurately decoded from its own event name representation $\mathbf{h}_{L,q}(x)$ (blue), reaching an MAE of roughly $3.5$--$4$\,s at intermediate layers. 
Predicting the unmentioned event's time $t_o$ from the same representation (orange) or the queried event's time from the first query token $\mathbf{h}_{L,1}(x)$ (green) yields consistently higher errors. 
The peak in decodability overlaps with the modality integration region identified by activation swapping, indicating that intermediate event name representations contain a coarse, event-specific temporal representation that generalizes across sound classes.

% % ============================================================================
\section{Geometry of Event-Bound Temporal Encoding}
\label{sec:geometry}
% ============================================================================

We next extract \emph{temporal IDs}—shared representations of events at the same temporal position—and analyze their  geometry.

\subsection{Extracting Temporal IDs}

We derive these IDs from the model's hidden event name representations when given 30\,s synthetic recordings containing a single ESC-50 event, paired with the query ``Is there [query event]?''. 
Let $\mathbf{h}_{L,q}(x)$ denote the layer-$L$ representation of the queried event name token, and let $\boldsymbol{\mu}_{L,c}$ be its training set mean for class $c$. 
We group examples into $2.5$\,s bins $B_k$ based on the event's midpoint and define the temporal ID of bin $k$ as the mean class-centered activation,
\begin{equation}
\boldsymbol{\tau}_{L,k}
=
\mathbb{E}\!\left[
\mathbf{h}_{L,q}(x)-\boldsymbol{\mu}_{L,c_q}
\mid t_q\in B_k
\right].
\label{eq:temporalid}
\end{equation}
Class centering removes variation dependent on event identity, so the resulting IDs primarily capture variation associated with temporal position. 
Bins with insufficient support are discarded, and the retained IDs are stacked into $\mathbf{T}_L\in\mathbb{R}^{K\times d}$ for the geometric analysis below.

\subsection{Temporal IDs Lie on a Low-Dimensional Curved Trajectory}
\label{sec:idgeometry}

\begin{figure}[h]
    \centering
    \includegraphics[width=\columnwidth]
    {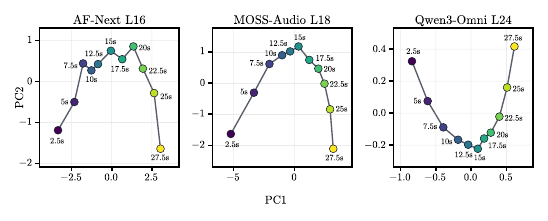}
    \caption{
Temporal ID geometry at representative modality integration layers. Points show temporal IDs projected onto the first two PCs.% Across models, the first two PCA-dimension capture more than $80\%$ of the position-dependent variance and the projected points form a systematic curved trajectory through activation space.
    }
    \label{fig:geometry}
\end{figure}

To characterize their geometry, we perform principal component analysis on $\mathbf{T}_L$ at representative modality integration layers identified in Section~\ref{sec:swap}.
Figure~\ref{fig:geometry} shows the resulting geometry. 
Temporal variation is strongly concentrated in a low-dimensional subspace: the first two principal components explain $83.9\%$, $86.4\%$, and $86.5\%$ of temporal ID variance for AF-Next, MOSS-Audio, and Qwen3-Omni, respectively. 
As event time increases, the temporal IDs trace a curved trajectory through this subspace.
Across models, early and late temporal IDs are strongly anti-aligned, with endpoint cosine similarities between $-0.71$ and $-0.73$, while ID magnitude decreases toward the recording midpoint and increases again toward the boundaries. 
This is consistent with a signed early-to-late component that passes close to the class-centered origin, together with weak off-axis variation producing the observed curvature.

Because the temporal IDs are extracted from fixed 30\,s recordings, it is unclear whether they represent absolute time or position relative to the recording duration. We test this using variable-duration RealDESED recordings. We class-center the event name activations, project them onto the first two PCs of the synthetic temporal IDs, and fit their projections as a function of either absolute midpoint $t_q$ or relative position $r_q=t_q/T_x$. Relative position consistently provides the better fit: multivariate $R^2$ increases from $0.19$ to $0.43$ for AF-Next, $0.31$ to $0.53$ for MOSS-Audio, and $0.30$ to $0.52$ for Qwen3-Omni. % Thus, the temporal ID trajectory primarily reflects position relative to recording duration rather than elapsed time in seconds.

\section{
Event-Bound Temporal Encoding Causally Mediates Coarse Reasoning
}
\label{sec:steering}
% ============================================================================

We next test whether temporal IDs affect temporal reasoning by steering activations and measuring prediction shifts.

\subsection{Temporal ID Steering}
\label{sec}

For each model, we reconstruct the temporal ID trajectory from the fixed-length synthetic training data at the modality integration layer selected above. 
For a queried event occurring at relative time $r_q=t_q/T$, we interpolate this trajectory to obtain its temporal ID $\boldsymbol{\tau}_L(r_q)$.  We then define an example-specific steering direction from the queried event's current position toward either endpoint,
\begin{equation}
\mathbf{d}_L(x)
=
\frac{
\boldsymbol{\tau}_L^{\mathrm{target}}
-
\boldsymbol{\tau}_L(r_q)
}{
\left\| \boldsymbol{\tau}_{L,1} - \boldsymbol{\tau}_{L,K} \right\|_2
},
\label{dir}
\end{equation}
where the target is the beginning or end of the synthetic temporal ID trajectory. Forward intervention uses target $\boldsymbol{\tau}_{L,K}$ to steer the belief of event occurrence to the end of the recording, and a backward intervention uses target $\boldsymbol{\tau}_{L,1}$ to move it toward the beginning. During inference, we add this direction to the query's event name tokens,
\begin{equation}
\widetilde{\mathbf{h}}_{L,i}(x)
=
\mathbf{h}_{L,i}(x)
+
\alpha\,\rho_L\,\mathbf{d}_L(x),
\label{eq}
\end{equation}
where $\alpha$ controls intervention strength and $\rho_L$ is the norm of the selected token activations.
% Thus, unlike a single global early-to-late direction, the intervention follows the discovered temporal ID trajectory from each queried event's current relative position toward the selected endpoint. The trajectory is estimated entirely from synthetic audio and applied without modification to natural RealDESED recordings.

\subsection{Steering Before/After Beliefs}

\begin{figure}[t]
\centering
\includegraphics[width=\columnwidth]{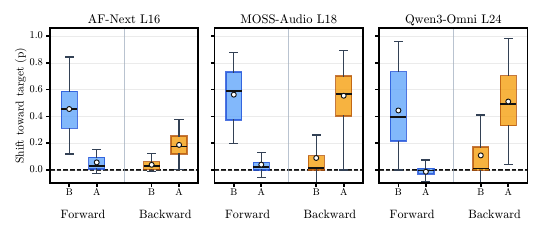}

\caption{Relation steering. Before (B) and After (A) denote groups which were predicted as before/after prior to the intervention. Steering shifts probability toward the target relation.
}
\label{fig:steering}

\end{figure}

We first evaluate queries of the form ``Does [query event] occur before or after [reference event]?''. We use teacher forcing with prefix ``[query event] occurs'' and measure the change in probability distribution over next tokens.
Figure~\ref{fig:steering} shows that steering shifts probability mass toward the intended relation across all three models: forward steering increases $p(\mathrm{after})$, whereas backward steering increases $p(\mathrm{before})$. Examples already aligned with the intervention---predicted as ``after'' before forward steering or ``before'' before backward steering---show only moderate probability shifts, as their predictions are often already saturated. Their original before/after predictions are preserved in 99.5\% of cases. In contrast, predictions opposing the intervention shift strongly toward the intended relation and are frequently flipped, with flip rates of 82\% for AF-Next, 84.2\% for MOSS-Audio, and 70.7\% for Qwen3-Omni. Thus, moving the queried event along the temporal ID trajectory consistently changes both probability mass and discrete before/after judgments on natural audio.

% % ============================================================================
\subsection{Steering Precise Sound Event Detection}
\label{sec:sed}
% ============================================================================

We next test whether temporal IDs also mediate precise onset prediction on a subset of RealDESED. To this end we select recordings in which the queried event class occurs exactly once. Baseline onset detection MAEs are $2.01$\,s for MOSS-Audio, $2.21$\,s for Qwen3-Omni, and $6.48$\,s for AF-Next. AF-Next is strongly biased toward a few onset values: 56.1\% and 13.8\% of predictions are $0$,s and $1.5$,s. We apply the same endpoint-directed intervention and compare predicted onsets before and after steering. 

\begin{figure}[t]
    \centering
    \includegraphics[width=\columnwidth]{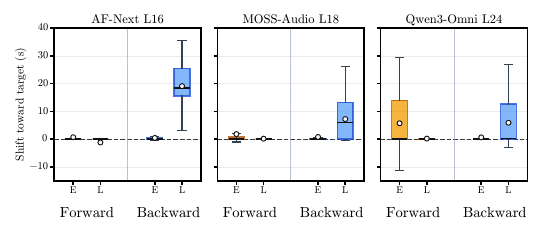}
\caption{
Onset steering. Early (E) and Late (L) examples were predicted as starting in the first and second half of the recording before the intervention.
}
    \label{fig:onset_steering}
\end{figure}

As shown in Figure~\ref{fig:onset_steering}, steering does not induce a consistent timestep displacement. AF-Next and MOSS-Audio respond asymmetrically: forward steering has little effect, whereas backward steering can substantially shift late predictions. For AF-Next, the backward intervention largely resets predictions to $0$\,s. Qwen3-Omni shows weaker but more bidirectional shifts toward the target. These asymmetric, model-dependent effects suggest that temporal IDs encode a coarse position which is only able to bias or override a separate fine-grained detection mechanism.

\section{Discussion and Conclusion}

Our activation swapping experiments showed that across the three investigated LALMs, event-specific temporal information becomes bound to event name representations in intermediate modality integration layers.
These representations appear to encode event position on a low-dimensional trajectory that reflects relative position of the event within the recording. 
% This structure is identified using controlled synthetic audio but generalizes to natural, variable-duration recordings, suggesting that it captures a representation used beyond the synthetic setting. 
Our activation swapping and before/ after steering results provide evidence that this temporal information is involved in temporal reasoning.
However, the temporal IDs do not appear to encode precise event timestamps: We find that nonlinear decoding predicts event midpoints with an MAE of around 3.5--4\,s, which is too coarse for precise localization. Furhtermore, steering with these IDs does not induce consistent shifts in predicted event onsets. Together, these results suggest that LALMs use event-bound temporal IDs primarily for coarse temporal reasoning, while relying on additional mechanisms for precise event localization.

Our analysis is limited to three models, a small set of query formulations, and relatively simple acoustic scenes with isolated query events and without strong overlaps. Future work should extend this analysis to more challenging acoustic conditions and investigate the mechanisms underlying fine-grained temporal localization.

\section{Acknowledgment}
\label{sec:ack}
The LIT AI Lab is supported by the Federal State of Upper Austria. GPT-5.6 Sol was used to assist with language editing and with writing code for the experiments in this manuscript. All AI-generated text and code was reviewed, verified, and revised by the authors.

% ==========================================================================
% References
% ==========================================================================

\bibliographystyle{IEEEbib}
\bibliography{strings,refs}

\end{document}